\documentclass[a4paper,fleqn]{cas-dc}
\usepackage{float}
\usepackage{hyperref}
\definecolor{mylightblue}{RGB}{0,128,172} 
\hypersetup{
    colorlinks=true,
    citecolor=mylightblue, 
    linkcolor=mylightblue, 
    urlcolor=mylightblue 
}

\usepackage{listings}
\usepackage{xcolor} 
\usepackage{caption}
\definecolor{deepblue}{rgb}{0,0,0.5}
\definecolor{lightblue}{rgb}{0.5,0.5,1}
\definecolor{mygreen}{rgb}{0,0.6,0}

\usepackage[numbers,sort&compress]{natbib}

\def\tsc#1{\csdef{#1}{\textsc{\lowercase{#1}}\xspace}}
\tsc{WGM}
\tsc{QE}
\begin{document}
\let\WriteBookmarks\relax
\def\floatpagepagefraction{1}
\def\textpagefraction{.001}
\let\printorcid\relax
\shorttitle{}    
\shortauthors{}  
\title [mode = title]{CrossInspector: A Static Analysis Approach for Cross-Contract Vulnerability Detection}  
\author[1]{\textcolor{black}{Xiao Chen}}
\ead{chenx553@mail2.sysu.edu.cn} 


\address[1]{School of Software Engineering, Sun Yat-Sen University, Guangzhou 510275, China}

\cortext[1]{Corresponding author} 

\begin{abstract}
With the development of blockchain technology, the detection of smart contract vulnerabilities is increasingly emphasized. However, when detecting vulnerabilities in inter-contract interactions (i.e., cross-contract vulnerabilities) using smart contract bytecode, existing tools often produce many false positives and false negatives due to insufficient recovery of semantic information and inadequate consideration of contract dependencies.

We present CrossInspector, a novel framework for detecting cross-contract vulnerabilities at the bytecode level through static analysis. CrossInspector utilizes a trained Transformer model to recover semantic information and considers control flow, data flow, and dependencies related to smart contract state variables to construct a state dependency graph for fine-grained inter-procedural analysis. Additionally, CrossInspector incorporates a pruning method and two parallel optimization mechanisms to accelerate the vulnerability detection process.

Experiments on our manually constructed dataset demonstrate that CrossInspector outperforms the state-of-the-art tools in both precision (97\%) and recall (96.75\%), while also significantly reducing the overall time from 16.34 seconds to 7.83 seconds, almost on par with the fastest tool that utilizes bytecode for detection. Additionally, we ran CrossInspector on a randomly selected set of 300 real-world smart contracts and identified 11 cross-contract vulnerabilities that were missed by prior tools.

\end{abstract}



\begin{keywords}
 Blockchain\sep Smart contract \sep Vulnerability detection \sep Static analysis\sep
\end{keywords}

\maketitle

\section{Introduction}\label{s1}

\begin{figure}
    \centering
    \includegraphics[width=.36\textwidth]{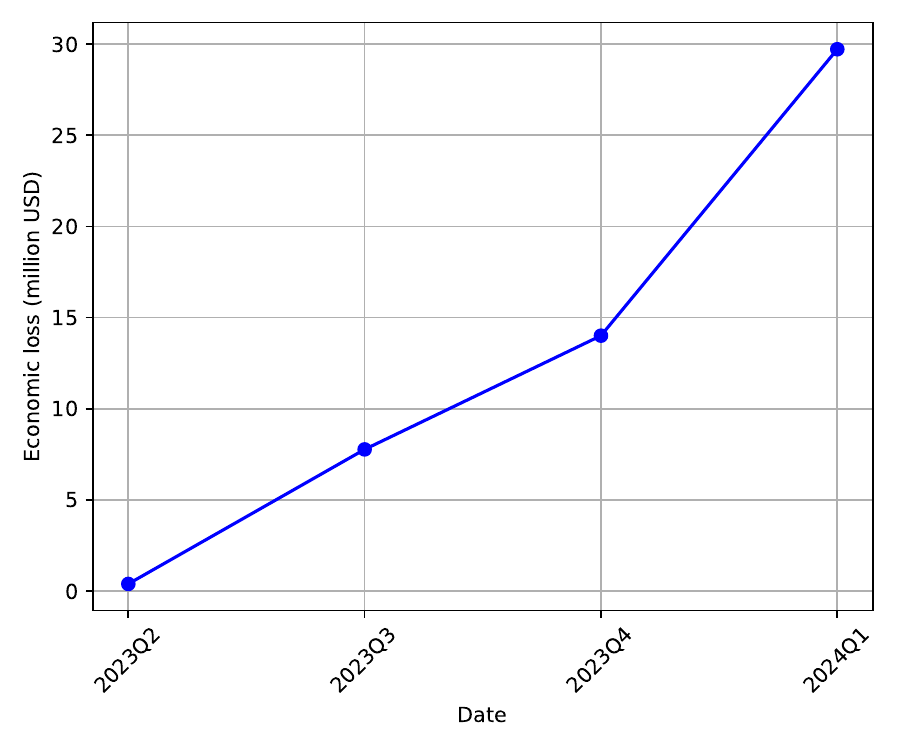}
    \caption{Economic loss caused by cross-contract vulnerabilities in the past year.}
    \label{跨合约漏洞统计}
\end{figure}

Smart contracts\cite{smartcontract} are computer programs stored on a blockchain\cite{blockchainoverview}, typically comprising a set of functions and state variables. Ethereum\cite{ethereum} is currently one of the most popular smart contract platforms. Ethereum hosts over 61 million smart contracts\cite{dune}, with approximately 99\% of them not open-sourcing their source code but existing in bytecode form. Smart contracts carry a large amount of digital assets, attracting hackers to exploit vulnerabilities for profit. Differing from vulnerabilities within a single contract or function, attackers nowadays tend to exploit vulnerabilities existing in interactions among multiple contracts (i.e., cross-contract vulnerabilities)\cite{smartdagger} to launch attacks. As illustrated in \hyperref[跨合约漏洞统计]{Fig. \ref*{跨合约漏洞统计}}, in the past year, the cumulative economic losses caused by cross-contract vulnerabilities have reached \$30 million\cite{slomist}.

Researchers have previously developed various tools to detect smart contract vulnerabilities (such as Mythril\cite{mythril}, Oyente\cite{oyente}, and Slither\cite{slither}). However, most tools primarily target vulnerabilities within a single contract, and their effectiveness in detecting cross-contract vulnerabilities is limited because they do not perform cross-contract inter-procedural analysis. Although the latest work, SmartDagger\cite{smartdagger}, can detect cross-contract vulnerabilities through bytecode, it does not recover sufficient semantic information or adequately consider contract dependencies. 

Unlike traditional programs, smart contracts typically involve state variables\cite{solidity} such as balance and address. However, the process of compiling them into bytecode often results in the loss of semantic information about these variables. Without sufficient semantic information, it may be challenging to accurately understand the interaction behavior of the contract, leading to false positives. Moreover, changes in smart contract state variables can affect the execution logic of functions, playing a crucial role in key operations such as fund management and permission control. Different functions may have read-write dependencies on the same state variable, and inadequate analysis of these dependencies may result in missed vulnerabilities. Additionally, detecting cross-contract vulnerabilities may encounter the problem of path explosion, where some tools may take tens of minutes to detect vulnerabilities in large contracts spanning thousands of lines of code. Therefore, there still lacks a method that is accurate and efficient enough to detect cross-contract vulnerabilities.

In this paper, we introduce CrossInspector, a new framework based on static analysis for detecting cross-contract vulnerabilities at the bytecode level. We collected over twenty thousand smart contract functions, categorized and described common semantic of state variables. Subsequently, we manually annotated the corresponding semantic information of state variables in the decompiled code and trained a Transformer-based\cite{attention} model to recover contract semantics. We integrate control flow dependencies, data flow dependencies, and state dependencies to construct a state dependency graph, modeling the interaction behavior and dependencies among multiple smart contracts, enabling fine-grained inter-procedural analysis. To quickly identify entry paths for vulnerabilities, we first propose a method based on pruning the inter-function call graph to remove irrelevant functions related to vulnerabilities. We then design a parallel memoization search\cite{mem} method, employing multi-process parallel search and sharing search results among processes to avoid redundant path traversal. Upon finding paths, we introduce a parallel taint analysis method, conducting multi-process data flow analysis on the state dependency graph to quickly identify affected functions and state variables.

The main contributions of this work are as follows:
\begin{itemize}
\item We propose a Transformer-based method to recover semantic information of state variables in smart contracts, and extract control flow, data flow, and state dependencies to construct a state dependency graph for detecting four common vulnerabilities, thereby enhancing the effectiveness of cross-contract vulnerability detection.
\item We introduce a pruning method and two parallel optimization mechanisms (namely, vulnerability entry path finding based on parallel memoization search and parallel taint analysis) to improve the efficiency of cross-contract vulnerability detection.
\item We conducted experiments on a manually constructed dataset and compared it with state-of-the-art tools. We found that CrossInspector not only surpasses other tools in effectiveness but also achieves comparable time efficiency to the fastest tool that utilizes bytecode for detection.
\end{itemize}

The rest of this paper is organized as follows: Section \ref{s2} introduces the relevant background knowledge. Section \ref{s3} describes the workflow of CrossInspector. Section \ref{s4} details the mechanisms within CrossInspector. Section \ref{s5} conducts experimental evaluations. Section \ref{s6} discusses related work. Section \ref{s7} summarizes the work of this paper.





\section{Background and motivation}\label{s2}

\subsection{Ethereum and smart contract}

Ethereum\cite{ethereum} is a decentralized platform that supports smart contracts\cite{smartcontract}, allowing developers to build decentralized applications\cite{decentralizedapplication}. Smart contracts are typically written in Solidity\cite{solidity}, consisting of a set of functions and state variables, and are compiled into bytecode executed by the Ethereum Virtual Machine (EVM). With the increasingly widespread application of smart contracts in finance, gaming, social, and other fields, the value of assets held by smart contracts is also growing, attracting many hackers attempting to exploit vulnerabilities for profit. For example, on March 28, 2024, the decentralized lending protocol Prisma Finance on Ethereum suffered a hack, resulting in a loss of \$11 million\cite{slomist}.

\subsection{Cross-contract vulnerability}

\hyperref[智能合约间调用示例]{Fig. \ref*{智能合约间调用示例}} illustrates a simple auction system containing a cross-contract timestamp manipulation\cite{timestamp} vulnerability. The \textit{block.timestamp} represents the timestamp of the block, i.e., the time when the block is packed, and relying on it for critical decisions can lead to security problems. The vulnerability arises when the \textit{Auction.bid} calls \textit{FundsHandler.recordBid}, and the function \textit{FundsHandler.recordBid} internally depends on \textit{block.timestamp} to determine if the auction has ended. Specifically, an attacker can send a bid transaction near the auction's end time (\textit{endTimestamp}). Since miners have the capability to make slight adjustments to the timestamp of the block containing the transaction, the attacker can (in collusion with the miners) have their bid accepted at the last moment of the auction, even though the bid is actually submitted after the auction ended. Consequently, other participants have almost no chance to react before the auction ends, leading to the attacker winning the auction at the last moment.

\begin{figure}[h!]
  \centering
  \begin{lstlisting}
contract FundsHandler {...
    address public auction;
    address public seller;
    address public itemOwner;
    mapping(address => uint) public refunds;
    uint public fee;
    uint public endTimestamp;
    function recordBid(address bidder) public payable onlyAuction {
        require(block.timestamp <= endTimestamp, "Cannot bid, auction ended");
        refunds[bidder] += msg.value;}
    function finalizeAuction(address highestBidder) public onlyAuction {
        require(block.timestamp > endTimestamp);
        require(refunds[highestBidder] > fee, "No valid bids or funds to transfer");
        seller.call.value(refunds[highestBidder] - fee)();
        itemOwner = highestBidder;}
...}
contract Auction {...
    address public owner;
    address public highestBidder;
    uint public highestBid;
    FundsHandler public fundsHandler;
    address[] public bidders;
    function bid() public payable {
        fundsHandler.recordBid.value(msg.value)(msg.sender);
        if (msg.value > highestBid) {
            highestBidder = msg.sender;
            highestBid = msg.value;}
        bidders.push(msg.sender);}
    function endAuction() public {
        require(msg.sender == owner, "Only the owner can end the auction");
        fundsHandler.finalizeAuction(highestBidder);}
...}
\end{lstlisting}
\caption{An example of cross-contract timestamp manipulation.}
    \label{智能合约间调用示例}
\end{figure}

Although \textit{FundsHandler.recordBid} is restricted with \textit{onlyAuction}, meaning only the \textit{Auction} contract can call this function, making it impossible to directly trigger this timestamp vulnerability, there exists a path \textit{Auction.bid} → \textit{FundsHandler.recordBid}, which indirectly leads to the occurrence of the vulnerability. Furthermore, it is apparent that this vulnerability actually affects the owner of the auction item, \textit{itemOwner}, which needs to be determined in the state dependency analysis in Section \ref{s42}. Therefore, when conducting cross-contract vulnerability detection, it is essential to thoroughly analyze the invocations and dependencies among multiple contracts.

\subsection{Prior research and their limitations}

Most vulnerability detection tools typically target vulnerabilities within a single contract. Due to the lack of cross-contract inter-procedural analysis, there are often many false positives and negatives when detecting cross-contract vulnerabilities. In cross-contract vulnerability detection tools that support bytecode as input, Pluto\cite{pluto} constructs the inter-contract control flow graph and employs symbolic execution to determine the existence of vulnerabilities. Smartdagg-er\cite{smartdagger} trains a Neural Machine Translation (NMT)\cite{nmt} model to recover contract attribute information, then constructs the cross-contract control flow graph and detects vulnerabilities through pruning and data flow reuse in taint analysis. They mainly suffer from three shortcomings.

Firstly, they do not effectively recover enough semantic information. Smartdagger recovers too few contract attributes from bytecode to cover common contract semantics. Additionally, the Long Short-Term Memory (LSTM) \cite{lstm} and the Graph Gated Neural Network (GGNN) \cite{ggnn} it employs encounter performance bottlenecks when dealing with large-scale or highly complex code. Without sufficient semantic information, tools may fail to accurately understand the behavior of contracts, leading to false positives.

Secondly, they do not fully consider contract dependencies. Pluto does not consider data flow dependencies among contracts, which may result in some unreachable paths being falsely reported as vulnerabilities. Smartdagger does not consider contract state dependencies, resulting in an incomplete analysis of paths affected by vulnerabilities.

Furthermore, the detection of cross-contract vulnerabilities may face the issue of path explosion, where some tools may take tens of minutes to detect vulnerabilities in large contracts spanning thousands of lines.

\section{Design of CrossInspector}\label{s3}
\subsection{Challenges and Solutions}



In this section, we clarify the challenges faced in current cross-contract vulnerability detection and introduce our solutions.

\textbf{Challenge 1: Semantic Recovery.}
State variables are widely used in inter-contract interactions, but decompilers usually represent state variables simply as \textit{stor\_0}, \textit{stor\_1}, and so on, requiring the recovery of their semantics. Previous work, such as SmartDagger\cite{smartdagger}, can only recover a few contract properties from bytecode, which is insufficient to cover common contract semantics. Additionally, the LSTM and GGNN-based semantic recovery model used by SmartDagger encounters performance bottlenecks, i.e., poor accuracy, when the code size is large or complexity is high.

To address this challenge, we collected smart contract function source code and classified the common semantics of state variables. We manually annotated the corresponding state variable semantics in the decompiled code to build a decompiled-level smart contract corpus. Subsequently, we proposed a Transformer-based semantic recovery model for smart contracts, using this model to recover contract state variable semantics from decompiled code, thereby enhancing understanding of contract behavior and reduce false positives.

\textbf{Challenge 2: Dependency Modeling.}
We need to conduct cross-contract dependency analysis to model contract interaction processes. Most smart contract analysis tools typically consider control flow and data flow dependencies while ignoring state dependencies, resulting in incomplete analysis and false negatives. Although SmartState\cite{smartstate} considers state dependencies, its dependency definition at the function level lacks granularity.

To address this challenge, we comprehensively consider control flow dependencies, data flow dependencies, state read-write dependencies, and state revert dependencies in smart contracts. We propose a method to extract fine-grained state dependencies at the basic block level. Based on the given smart contract bytecode, we extract and analyze dependencies, construct a state dependency graph, and apply it to the detection of four common vulnerabilities, thereby identifying a broader scope of vulnerable areas.

\textbf{Challenge 3: Efficiency Optimization.}
Cross-contract call paths have many branches, which may lead to path explosion and high time overhead. Specifically, previous cross-contract vulnerability detection work can only detect one vulnerability at a time, requiring a repeat of the detection process when a new vulnerability is detected.

To address this challenge, we propose two parallel optimization mechanisms to accelerate the vulnerability entry path search and taint analysis processes. The parallel memoization search technique targets the multi-source multi-target problem, allowing the search for paths from multiple function entry points to multiple vulnerability indicators simultaneously, sharing search results among different processes to reduce redundant traversals. In the parallel taint analysis technique, each process handles a portion of the data flow taint propagation, merging states at the end of all processes, effectively improving the speed of taint analysis.

\subsection{Workflow of CrossInspector}
\begin{figure*}
    \centering
    \includegraphics[width=.9\textwidth]{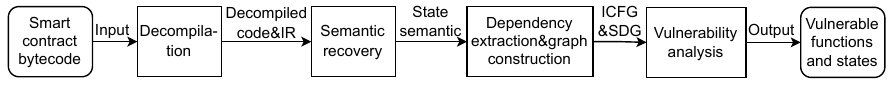}
    \caption{The workflow of CrossInspector.}
    \label{overview}
\end{figure*}

The workflow of CrossInspector is illustrated in \hyperref[overview]{Fig. \ref*{overview}}. Initially, the bytecode of smart contracts is decompiled using the existing decompiler (Gigahorse\cite{gigahorse}) to obtain the decompiled code and the intermediate representation (IR) in three-address form. Subsequently, semantic information of state variables in the decompiled code is recovered (see Section \ref{s41}). Then, dependencies are extracted from the IR to sequentially construct the inter-contract control flow graph (ICFG) and the state dependency graph (SDG) (see Section \ref{s42}). Finally, vulnerability detection based on taint analysis is performed on the constructed ICFG and SDG to identify functions and state variables that may be affected by vulnerabilities (see Section \ref{s43}).

\subsection{Vulnerability indicators}

The \hyperref[漏洞类型及indicator识别规则]{Table \ref*{漏洞类型及indicator识别规则}} presents four types of vulnerabilities addressed in this paper along with their corresponding indicator identification rules, namely reentrancy\cite{reentrancy}, timestamp manipulation\cite{timestamp}, denial-of-service (DoS) attack\cite{dos1,dos2}, and integer overflow\cite{integerflow}. The detection rules for these indicators are derived from our analysis of the opcode characteristics of vulnerabilities, similar but slightly different from the detection rules in previous works.Although this paper considers only four types of vulnerabilities, CrossInspector is a generic framework that can cover other types of vulnerabilities by integrating more detection rules.

\textbf{Reentrancy.} Involves external calls and uses call.value for transferring funds.

\textbf{Timestamp manipulation.} Opcode related to timestamps, i.e., block.timestamp, block.number, or now, appears in conditional statements.

\textbf{DoS.} Involves external calls within loops.

\textbf{Integer Overflow.} Uses addition, multiplication, or division without overflow checks. 

Although smart contracts from Solidity version 0.8.0 onwards can avoid integer overflows through built-in SafeMath overflow checks, developers sometimes temporarily disable integer overflow checks in code blocks using the unchecked keyword to save Gas. Therefore, detecting integer overflow remains necessary.

\begin{table}[pos=h!]
    \centering
    \footnotesize
    \caption{Vulnerability indicator rules.}
    \label{漏洞类型及indicator识别规则}
    \begin{tabular}{|l|m{4cm}|}
    \hline
    Vulnerability type & Vulnerability indicator rule \\
    \hline 
    Reentrancy & isContained(Externalcall)\(\land\)
    isContained(call.value) \\
    \hline
    Timestamp manipulation & isExecutionLock(Timestamp) \\
    \hline
    DoS attack & isInLoop(ExternalCall) \\
    \hline
    Integer overflow & isContained(Add\(\lor\)Sub\(\lor\)Mul)\(\land\)
    (\(\neg\)isChecked(Overflow)) \\
    \hline
    \end{tabular}
\end{table}
\section{Approach details}\label{s4}
\subsection{Semantic recovery based on Transformer}\label{s41}
In this section, we introduce a semantic recovery method based on Transformer.
\subsubsection{Model design}

We employ Transformer\cite{attention} for the task of semantic recovery in smart contracts. The choice of this model is primarily based on its efficient training speed and excellent long sequence processing capabilities, which contribute to recovering high-quality semantic information.

\begin{figure}[h!]
    \centering
    \includegraphics[width=.45\textwidth]{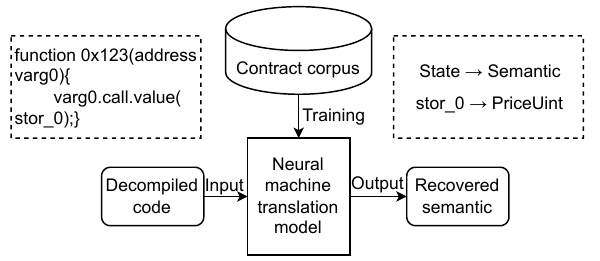}
    \caption{Working process of semantic recovery.}
    \label{语义恢复流程}
\end{figure}

We have designed a Transformer-based semantic recovery model, the overall process of which is illustrated in \hyperref[语义恢复流程]{Fig. \ref*{语义恢复流程}}. The entire process begins with inputting the bytecode of the target smart contract into a decompiler to generate decompiled code. Next, functions containing state variables are extracted and transformed into token streams. Finally, these token streams are fed into the semantic recovery model, which, after training, is able to generate the corresponding semantic information for each state variable.

\begin{figure}[h!]
    \centering
    \includegraphics[width=.4\textwidth]{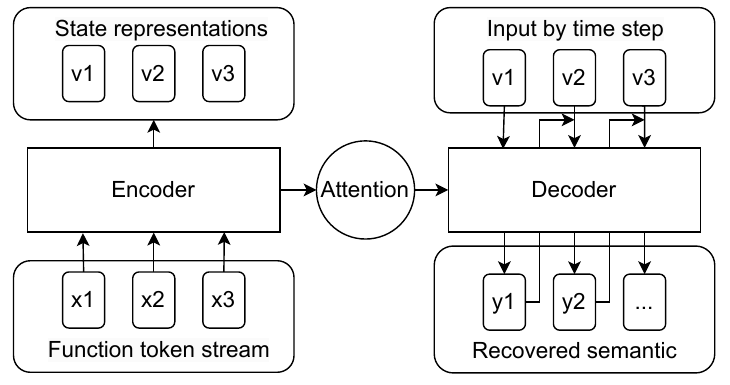}
    \caption{Details of encoder and decoder.}
    \label{语义恢复模型细节}
\end{figure}
\hyperref[语义恢复模型细节]{Fig. \ref*{语义恢复模型细节}} provides an overview of the model structure, which follows an encoder-decoder architecture. The encoder is responsible for transforming the function token stream, extracting state variables and rich contextual information, while the decoder uses this information to predict the specific semantic of each state variable. Below we detail the working principles of the encoder and decoder.

\textbf{Encoder.}
The encoder transforms the input function token stream $X = (x_1, x_2, \dots, x_n)$ into a set of representations, denoted as $H = (h_1, h_2, \dots, h_n)$. During the training process, the encoder effectively encodes information relevant to each input token $x_i$ into its corresponding representation $h_i$ through learning.
In function code, a state variable may occur multiple times. We perform pooling operations on the representations $h_i$ of each occurrence position of the state variable, generating a consolidated representation $v$ to more accurately reflect the overall semantic information of that state variable.

\textbf{Decoder.}
At each timestep, the decoder receives the output from the encoder $H = (h_1, h_2, \dots, h_n)$ and the representation of each state variable $v_t$ as input. Moreover, when predicting the semantic $y_t$ of the state variable $v_t$, it also takes all previously predicted semantics from $y_1$ to $y_{t-1}$ as input. Thus, the decoding process can be represented by the function $z_t = f_{d}(y_1, \ldots, y_{t-1}, v_t, H; \theta_{d})$, where $\theta_d$ is the parameters of the decoder. The output layer of the decoder computes a temporary output $s_t = Wz_t + b$, where $W$ and $b$ are learned weights and biases, respectively. By applying the softmax function, $s_t$ is transformed into a normalized probability distribution, i.e., $P_r(y_t \mid y_1, y_2, \dots, y_{t-1}, X) = \text{softmax}\ s_t$. This step aims to predict the conditional probability distribution of the next semantic $y_t$ given the previous $t-1$ semantics. If there are $n$ state variable semantics to predict, then the overall process probability can be expressed as
$P_r(Y \mid X) = \prod\limits_{t=1}^n P_r(y_t \mid y_1, y_2, \ldots, y_{t-1}, X)
$.

We adopt the Beam Search \cite{beamsearch} algorithm, which maintains computational efficiency while more comprehensively exploring the solution space, thereby increasing the probability of finding the optimal solution.

\subsubsection{Semantics classification and data collection}
In this section we first categorize the semantic information of state variables. Through the collaborative efforts of two domain experts, a series of detailed semantic categories have been constructed to cover the most common uses of state variables in smart contracts. These semantic categories include but are not limited to the contents in \hyperref[语义表]{Table \ref*{语义表}}. For example, \textit{BalanceMapping} represents a state variable used to record balances in a mapping type.

\begin{table}[pos=h!]
\caption{Semantic categories and descriptions.}
\label{语义表}
\centering
\footnotesize
\begin{tabular}{ll}
\toprule
Semantic category & description \\
\midrule
AmountUint       &  quantity \\
TimeUint         & timestamps \\
PriceUint        & transaction amounts or fees \\
SupplyUint       & supply or inventory \\
NameString       & token or project names \\
SymbolString     & token or project symbol \\
UriString        & links pointing to external content \\
BalanceMapping   & account balances \\
AllowanceMapping & authorized limits\\
PausedBool       & whether a function is paused \\
EnableBool       & whether a function is enabled \\
NonreentrantBool & reentry state \\
OwnerAddress     & the owner's address \\
WalletAddress    & an external wallet address \\

\bottomrule
\end{tabular}
\end{table}

To construct a training dataset, we collected source code and bytecode of smart contracts from Ethereum. The bytecode was decompiled into decompiled code resembling source code. By observing and comparing the source code with the decompiled code, we manually annotated the semantic information of the decompiled code's state variables corresponding to their source code.

We collected over twenty thousand smart contract functions. The dataset was randomly partitioned into training, validation, and test sets based on contract names, with a ratio of 8:1:1. This file-based partitioning method ensures that information from the same smart contract does not simultaneously appear in different sets, thereby reducing the risk of data leakage. Additionally, we employed Byte-Pair Encoding (BPE)\cite{bpe} as a preprocessing step to reduce the size of the vocabulary and enhance the representation capability for rare words.

\subsection{State dependency analysis and graph construction}\label{s42}
In this section, we introduce the extraction of dependency information from smart contracts and the construction of the state dependency graph. In addition to control flow dependencies and data flow dependencies, we particularly focus on the state read-write dependencies and state revert dependencies in smart contracts.

\subsubsection{State dependency analysis}
State read-write dependencies describe the read and write operations on state variables in a smart contract. Specifically, when a basic block writes to a state variable s, a directed edge is added from that basic block in the smart contract control flow graph to the state variable s, indicating the flow of data output. Conversely, when a basic block reads \vspace{-3pt}a state variable s, an edge is added from the state variable s to that basic block, representing the source of input data.

In addition to state read-write dependencies, we also focus on state revert dependencies. This dependency arises from the unique way in which assertions and exceptions are handled in smart contracts. Several assertion and exception-related operations are introduced in smart contract design, namely \textit{assert}, \textit{require}, and \textit{revert}, which trigger state revert upon the failure of function execution. These operations underlyingly utilize the EVM's REVERT opcode to undo all state changes of the current transaction.

When a basic block writes to a state variable s, and another basic block containing \textit{assert}, \textit{require}, or \textit{revert} operation needs to read and decide whether to roll back based on that state variable s, then there exists a state revert dependency between these two basic blocks concerning s. In the control flow graph, the dependency of state revert is represented by adding a directed edge between the basic block of the write operation and the starting point of the straight branch where the basic block triggering revert from the read operation is located.

\begin{figure}[h!]
    \centering
    \includegraphics[width=.48\textwidth]{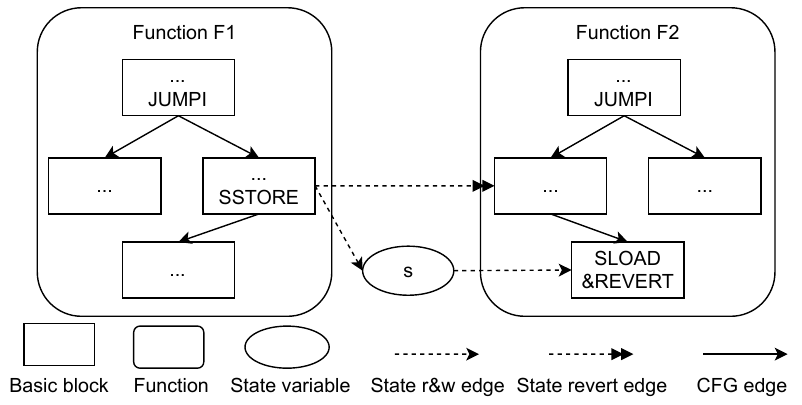}
    \caption{An example of state dependency.}
    \label{状态回退依赖示例}
\end{figure} 

In \hyperref[状态回退依赖示例]{Fig. \ref*{状态回退依赖示例}}, a basic block in function F1  writes to the state variable s, and a basic block in function  F2 reads the state variable s and uses it as an assertion condition. There exists a state revert dependency between these two basic blocks. The state revert dependency edge points to the starting point of the straight-line branch where the REVERT basic block is located, which is the previous basic block of the basic block containing \textit{SLOAD}, because operations within this branch will be rolled back when the condition for reading the state variable s is not satisfied, indicating that the write to s affects the entire branch.

\subsubsection{State dependency graph construction}
The Inter-Contract Control Flow Graph (ICFG) is used to depict inter-contract control flow dependencies. It is defined as a graph G=<N,E>, where N represents basic block nodes, and E consists of two types of edges: intra-procedural call edges and inter-procedural call edges.

\textbf{ICFG Construction.} Based on the three-address IR, basic blocks are connected by directed edges according to control flow instructions and jump targets to form CFG. Then, basic blocks in different CFGs are analyzed to identify each inter-procedural call operation, and inter-procedural call edges are added between callers and callees to obtain ICFG.

The State Dependency Graph (SDG) is defined as a graph G=<N,S,E>, where N represents basic block nodes, S represents state variable nodes, and E consists of three types of edges: control flow edges, state read-write dependency edges, and state revert dependency edges.

\textbf{SDG Construction.} Starting from the three-address IR of the smart contract, all state variable read-write operations are identified to obtain state read-write dependency edges. Then, analysis is conducted on basic blocks containing REVERT operations and the state variables with read dependencies on these basic blocks. And combined with state read-write dependency edges, state revert dependency edges are obtained. Finally, on the basis of ICFG, state variable nodes are added, along with state read-write dependency edges and state revert dependency edges, completing the construction of SDG.

\begin{figure}[h!]
    \centering
    \includegraphics[width=.45\textwidth]{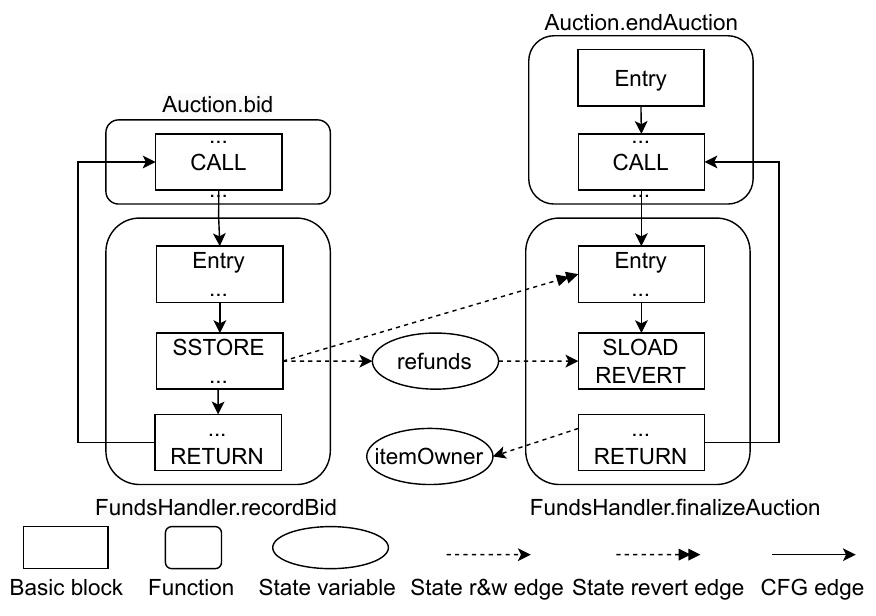}
    \caption{An example of state dependency graph.}
    \label{状态依赖图SDG示例}
\end{figure}

Taking the code in \hyperref[智能合约间调用示例]{Fig. \ref*{智能合约间调用示例}} as an example, we construct the SDG. For ease of viewing, most basic block and state variable nodes inside functions are omitted, as shown in \hyperref[状态依赖图SDG示例]{Fig. \ref*{状态依赖图SDG示例}}. This graph is composed of state variable nodes, state read-write dependency edges, and state revert dependency edges, based on the ICFG. The function \textit{FundsHandler.recordBid} writes to the state variable \textit{refunds}, and the function \textit{FundsHandler.finalizeAuction} reads the value of the state variable \textit{refunds} as a condition for \textit{require}. Therefore, a state revert dependency edge from the function \textit{FundsHandler.recordBid} points to the function \textit{FundsHandler.finalizeAuction} (here, for ease of viewing, points to the entry basic block, but actually points to the starting basic block of the linear branch). In the introduction of Section \ref{s2}, \textit{Auction.bid\hspace{0pt}→FundsHandler.recordBid} is the function call path for the timestamp manipulation. Due to the presence of state read-write dependency edges and state revert dependency edges, this path extends to the function \textit{FundsHandler.finalizeAuction}, and ultimately affects the state variable \textit{itemOwner}, meaning that this timestamp manipulation affects the final ownership of the auction item through dependency relations.

\subsection{Vulnerability detection based on parallel taint analysis}\label{s43}
In this section, we introduce efficiency optimization methods for vulnerability detection processes, including vulnerability entry path searching based on parallel memoization search, and leveraging parallel taint analysis techniques to determine the potential impact scope of vulnerabilities.

\subsubsection{Vulnerable entry trace finding based on parallel memoization search}
The number of functions in smart contracts can be substantial, with each function consisting of a series of basic blocks. Employing a brute-force search method would incur significant time consumption because this approach would explore numerous irrelevant paths, including those unrelated to vulnerability indicators.

We propose a pruning strategy based on the inter-function call graph. In the inter-function call graph, each node represents a function, and edges represent the calling relationships between functions. We focus only on the call relationships between functions, disregarding information about basic blocks within functions. The construction process of the inter-function call graph starts from a given list of functions and information about inter-procedural calls between basic blocks, adding edges between caller and callee functions based on inter-procedural call information. Then, all weakly connected subgraphs are computed and the set of functions containing entry basic blocks and the set of functions containing vulnerability indicators are filtered out. Next, for each weakly connected subgraph, we check if it simultaneously contains at least one function from both of these function sets. If so, all functions in that weakly connected subgraph are added to the set of functions to be analyzed. This approach yields all functions that may be related to vulnerabilities for further analysis.

Next, we search in the ICFG for feasible paths from entry basic blocks to vulnerability indicators. This is a multi-source multi-target search problem, as there may be multiple entry basic blocks and vulnerability indicators, and their paths may intertwine, leading to the need for repetitive traversal during analysis.

Parallel processing offers an effective way to accelerate the vulnerability detection process in smart contracts. For example, Park\cite{park} improves analysis efficiency through parallel accelerated symbolic execution. Specifically, whenever a path branching occurs, a new process is launched in parallel if there are available processor cores to handle it. This strategy proves to be effective in scenarios where symbolic execution involves extensive computations. However, in static analysis scenarios, adopting this strategy without restrictions may incur significant overhead due to frequent process spawning. Therefore, we employ a conservative parallel strategy by limiting the number of parallel processes to the number of entry basic blocks. This restriction prevents the unnecessary spawning of additional processes during the analysis, thereby controlling the overhead of process management while accelerating the search process.

Memoization search\cite{mem} is another key technology for improving path search efficiency. Specifically, by setting up a shared memory space for each parallel process to record traversed nodes and their subsequent path information, when any process encounters these traversed nodes again in subsequent search processes, relevant information can be directly obtained from the shared memory without the need for repeated traversal.

\begin{figure}[h!]
    \centering
    \includegraphics[width=.45\textwidth]{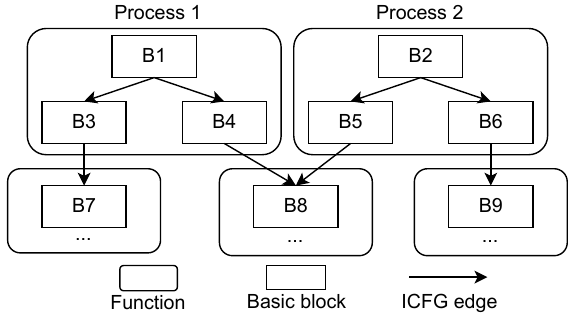}
    \caption{An example of parallel memoization path search.}
    \label{并行记忆化路径搜索示例}
\end{figure}

\hyperref[并行记忆化路径搜索示例]{Fig. \ref*{并行记忆化路径搜索示例}} depicts an example of an ICFG, with two entry basic blocks, B1 and B2. The common path for both processes begins at node B8. When traversing, two processes are initiated separately. Each process, upon traversal, writes the subsequent paths of the traversed nodes into shared storage. Consequently, following a depth-first order, process 1 completes the traversal of B7 and its subsequent paths, while process 2 completes the traversal of B8 and its subsequent paths. When process 1 reaches B8, it can directly access the traversal results of process 2 without the need for redundant traversal, significantly reducing time costs.

\subsubsection{Parallel taint analysis}

To perform taint analysis, it is necessary to identify the sources and sinks of taint first, and then propagate taint on the SDG to identify paths and variables that might be tainted. We considers several types of taint sources and sinks, as shown in \hyperref[污点源污点汇]{Table \ref*{污点源污点汇}}. Taint sources include parameters passed by contract callers and parameters of public functions. Taint sinks include external calls, state variables of smart contracts, and vulnerability indicators. The basic principle of taint propagation is based on data-flow analysis. Specifically, for basic block nodes in the SDG, if a right operand of an instruction is already in the tainted value list and its opcode is identified as a taint opcode, its left operand is added to the tainted value list. Conversely, if the right operand is not in the tainted value list or the opcode is not a taint opcode, this is recorded as a untainted path, meaning the left operand corresponding to it is recorded as its successor node.

\begin{table}[pos=h!]
\centering
\caption{Taint sources and taint sinks.}
\label{污点源污点汇}
\begin{tabular}{l|m{5cm}}
\hline
Taint source&Parameters of public functions, parameters passed by contract callers \\
\hline
Taint sink&State variables, vulnerability indicators, external calls \\
\hline
\end{tabular}
\end{table}

The core idea of parallel taint analysis strategy is to implement parallel taint analysis process on the SDG. Specifically, for each entry node, a separate process is initiated. In each process, the vulnerability entry path found in the previous section is traversed first to determine whether the taint can propagate to the vulnerability indicator (some types of vulnerabilities need to be confirmed by taint analysis, such as integer overflow). Then continue to traverse forward until the end. After all processes finish, taint state merging is performed. This step merges the tainted value lists identified in various processes into a global tainted value list. At the same time, previously recorded untainted paths are merged: if the right operand of a untainted path is added to the global tainted value list due to the analysis of other processes, the path is considered tainted, and the corresponding left operands are also added to the global tainted value list.

\begin{figure}[h!]
    \centering
    \includegraphics[width=.4\textwidth]{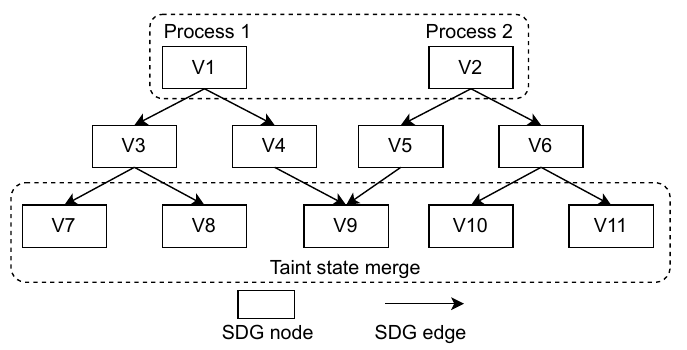}
    \caption{An example of parallel taint analysis.}
    \label{并行污点传播示例}
\end{figure}

\hyperref[并行污点传播示例]{Fig. \ref*{并行污点传播示例}} is an example of SDG. If taint propagation is performed in multiple processes starting from nodes V1 and V2 according to the taint propagation criteria, only two processes are initiated. Since the multiple processes are executed in parallel, if node V5 has tainted data flow from node V3, or if node V2 has tainted data flow from node V4, it needs to be discovered during the merging process after each inter-procedural analysis.

\begin{figure}[h!]
    \centering
    \includegraphics[width=.5\textwidth]{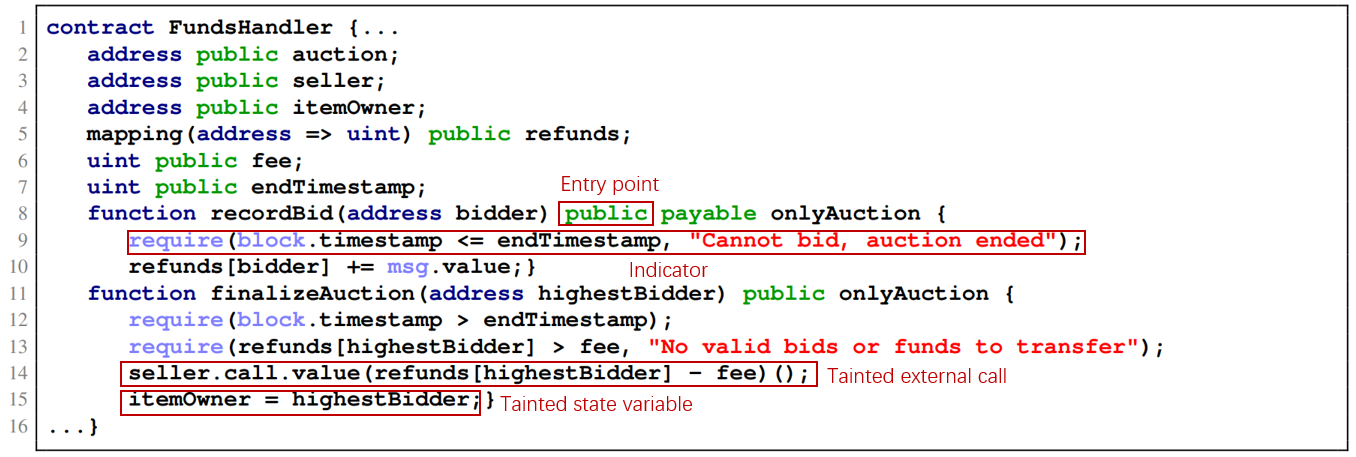}
    \caption{An example of vulnerability detection process.}
    \label{漏洞检测流程示例}
\end{figure}

Below, taking the contract \textit{FundsHandler} in \hyperref[智能合约间调用示例]{Fig. \ref*{智能合约间调用示例}} as an example, a simple vulnerability detection process using only serial mechanisms is demonstrated, as shown in \hyperref[漏洞检测流程示例]{Fig. \ref*{漏洞检测流程示例}}. First, the bytecode of the contract is decompiled, semantic recovery is performed, and the state dependency graph is constructed. Then, it is found that the function \textit{recordBid} contains a timestamp manipulation indicator. The path from the entry function to this indicator is searched. According to the analysis in Section \ref{s2}, it is known that this indicator can be entered by the function \textit{Auction.bid} of an external contract. After obtaining this feasible path, taint propagation starts from the entry function. According to the analysis in Section \ref{s3}, it is known that there is a state dependency relationship between the function \textit{recordBid} and the function \textit{finalizeAuction}. Taint propagates along the edges of the SDG, flows through the state dependency edges to the function \textit{finalizeAuction}, and ultimately taints the state variable \textit{itemOwner}. Therefore, through vulnerability detection, we confirms the existence of a cross-contract timestamp manipulation and reports the path \textit{Auction.bid\hspace{0pt}→FundsHandler.recordBid\hspace{0pt}→[refunds]\hspace{0pt}→FundsHandler.finalizeAuction\hspace{0pt}→[seller,itemOwner]}.

\section{Evaluation}\label{s5}



To evaluate the effectiveness and efficiency of the method proposed in this paper for detecting cross-contract vulnerabilities, we designed four research questions.

RQ1. How are the effectiveness and efficiency of CrossInspector in detecting cross-contract vulnerabilities compared to existing work?

RQ2. How do semantic recovery and dependency extraction analysis help in detecting cross-contract vulnerabilities?

RQ3. How effective are the parallel memoization path search and parallel taint analysis in reducing time overhead?

RQ4. Can CrossInspector detect cross-contract vulnerabilities in the real world?

The first dataset is a cross-contract vulnerability dataset $D_m$, which contains 334 smart contracts with cross-contract vulnerabilities. The raw data for this dataset is sourced from three previously published open-source GitHub repositories \cite{smartbugs,plutodataset,src3}. The repositories \cite{plutodataset,src3} are derived from two representative cross-contract vulnerability detection works. However, since the 200 contracts from these two repositories are not actual Ethereum contracts, they do not adequately reflect the diversity and complexity of real-world contracts. The repository \cite{smartbugs} compensates for this as it comes from a well-known empirical study \cite{icse20}, which includes 47,398 real-world Ethereum smart contracts. From this, we randomly selected 3,000 contracts. The vulnerabilities present in these 3,200 contracts were identified and annotated by two domain experts. We ultimately identified 334 smart contracts with cross-contract vulnerabilities, which include an additional 67 single-contract vulnerabilities.

The second dataset $D_r$ is sourced from the aforementioned SmartBugs repository \cite{smartbugs}. We retrieved the corresponding bytecode from Etherscan \cite{etherscan}. This dataset is used to further validate the practical applicability of the proposed method on real-world smart contracts. \hyperref[51]{Table \ref*{51}} provides information on these two datasets, including the number of smart contracts, types of vulnerabilities, and the number of vulnerabilities.


\begin{table}[pos=h!]
\centering
\footnotesize
\caption{Dataset vulnerability information.}
\label{51}
\setlength{\tabcolsep}{4pt} 
\begin{tabular}{lccccc}
\toprule
Dataset          & FileNum & Reentrancy & Timestamp. & DoS. & Overflow. \\ 
\midrule
$D_m$ & 334      & 103    & 157      & 73         & 68     \\
$D_r$       & 47,398     & -      & -        & -          & -      \\ 
\bottomrule
\end{tabular}
\end{table}

We aimed to select state-of-the-art open-source tools that cover the four types of vulnerabilities our study focuses on as comprehensively as possible. After comparison(See Section \ref{s6}), we selected four tools: Pluto\cite{pluto}, Mythril\cite{mythril}, Slither\cite{slither}, and Oyente\cite{oyente}.

All experiments were conducted on a server equipped with an Intel i9-10980XE CPU, an RTX3090 GPU, and 250 GB RAM running Ubuntu 20.04. Considering that the analysis process of some tools may take an excessively long time, we set a uniform timeout of 600 seconds.

\subsection{Overall effectiveness}

We compared CrossInspector with other tools on the dataset $D_m$ using two metrics: recall and precision. The precision and recall results are shown in \hyperref[精确率表]{Table \ref*{精确率表}} and \hyperref[召回率表]{Table \ref*{召回率表}}, respectively.

In terms of precision, CrossInspector achieved 97\%, which is higher than other tools. This difference indicates that other tools introduce many false positives, often due to a lack of in-depth semantic analysis. In terms of recall, CrossInspector reached 96.75\%, significantly higher than other tools. Their lower recall is mainly due to a lack of capability for cross-contract analysis or insufficient consideration of contract dependencies.

CrossInspector introduced 12 false positives. Upon manual inspection, most of these were found to stem from limitations in the decompiler. For instance, the Gigahorse decompiler sometimes fails to accurately infer constant constraints in statements, resulting in false positives when detecting integer overflow. 

CrossInspector missed 13 vulnerabilities. Upon manual inspection, most of these misses were found to originate from limitations in static analysis. For example, for timestamp manipulation, our rules are designed to search for timestamp-related opcodes and subsequent comparison instructions within the basic blocks, while in reality, these comparison instructions may appear in preceding basic blocks before the timestamp opcodes. In practice, cross-contract vulnerability scenarios are complex, and our aim is to provide detection rules that cover most cases rather than arbitrarily expanding rules based on the characteristics of missed vulnerabilities. This is actually a limitation of static analysis, as static analysis lacks the precision of running the program, unlike symbolic execution or formal verification.

\begin{table*}[h]
\centering
{\footnotesize
\setlength{\tabcolsep}{4pt} 
\caption{Precision of different tools on dataset $D_m$.}
\label{精确率表}
\begin{tabular}{l|lll|lll|lll|lll|lll}
\hline
Vulnerability & \multicolumn{3}{c|}{Reentrancy} & \multicolumn{3}{c|}{Timestamp.} & \multicolumn{3}{c|}{DoS attack} & \multicolumn{3}{c|}{Integer overflow} & \multicolumn{3}{c}{Overall} \\ \hline
    & TP & FP & prec. & TP & FP & prec. & TP & FP & prec. & TP & FP & prec. & TP & FP & prec. \\ \hline
Mythril & 91 & 136 & 40.08\% & 132 & 0 & 100.0\% & 15 & 139 & 9.74\% & 4 & 37 & 9.75\% & 242 & 312 & 43.68\% \\ 
Oyente & 36 & 12 & 75.0\% & 30 & 0 & 100.0\% & 2 & 6 & 25.0\% & 0 & 0 & 0& 68 & 18 & 79.06\% \\ 
Pluto & 98 & 13 & 88.28\% & 123 & 0 & 100.0\% & - & - & - & 50 & 0 & 100.0\% & 271 & 13 & 95.42\% \\
Slither & 69 & 37 & 65.09\% & 75 & 0 & 100.0\% & 24 & 3 & 88.88\% & - & - & - & 168 & 40 & 80.76\% \\ \hline
CrossInspector & 99 & 4 & 96.11\% & 153 & 0 & 100.0\% & 68 & 0 & 100.0\% & 68 & 8 & 89.47\% & 388 & 12 & 97\% \\ \hline
\end{tabular}
}
\end{table*}
\begin{table*}[h]
\centering
{\footnotesize
\setlength{\tabcolsep}{4pt} 
\caption{Recall of different tools on dataset $D_m$.}
\label{召回率表}
\begin{tabular}{l|lll|lll|lll|lll|lll}
\hline
Vulnerability & \multicolumn{3}{c|}{Reentrancy} & \multicolumn{3}{c|}{Timestamp.} & \multicolumn{3}{c|}{DoS attack} & \multicolumn{3}{c|}{Integer overflow} & \multicolumn{3}{c}{Overall} \\ \hline
    & TP & FN & Recall & TP & FN & Recall & TP & FN & Recall & TP & FN & Recall & TP & FN & Recall \\ \hline
Mythril & 91 & 12 & 88.34\% & 132 & 25 & 84.07\% & 15 & 58 & 20.54\% & 4 & 64 & 5.88\% & 242 & 159 & 60.34\% \\ 
Oyente & 36 & 67 & 34.95\% & 30 & 127 & 19.10\% & 2 & 71 & 2.73\% & 0 & 68 & 0 & 68 & 333 & 16.95\% \\ 
Pluto & 98 & 5 & 95.14\% & 123 & 34 & 78.34\% & - & - & - & 50 & 18 & 73.52\% & 271 & 57 & 82.6\% \\
Slither & 69 & 34 & 66.99\% & 75 & 82 & 47.77\% & 24 & 49 & 32.87\% & - & - & - & 168 & 165 & 50.45\% \\ \hline
CrossInspector & 99 & 4 & 96.11\% & 153 & 4 & 97.45\% & 68 & 5 & 93.15\% & 68 & 0 & 100.0\% & 388 & 13 & 96.75\% \\ \hline
\end{tabular}
}
\end{table*}

\subsection{Overall efficiency}

\hyperref[时间开销表]{Table \ref*{时间开销表}} lists the average detection time for each tool on dataset $D_m$. The average detection time for CrossInspector is 7.83 seconds, which is comparable to Oyente and Pluto but far better than Mythril, mainly because Mythril's symbolic execution analysis mechanism is more time-consuming.

\begin{table}[pos=h!]
\centering
\setlength{\tabcolsep}{3pt} 
\caption{Time cost of different tools on dataset $D_m$.}
\label{时间开销表}
\begin{tabular}{lccccc}
\hline
        Tool   & Mythril & Oyente & Slither & Pluto & CrossInspector \\ \hline
Avg.Time(s)    & 198.00       & 6.88      & 1.81       & 7.61     & 7.83       \\ \hline
\end{tabular}
\end{table}

Compared to Slither, CrossInspector's average detection time is slightly higher. This is mainly because Slither analyzes source code, while CrossInspector operates on bytecode, with an average of over 5 seconds used for the decompilation process.

\subsection{Ablation experiments}
To answer RQ2 and RQ3, we conduct separate ablation experiments for analysis.
\subsubsection{Semantic Recovery}


One major advantage of CrossInspector is semantic recovery, particularly significant in detecting integer overflow and reentrancy vulnerabilities. As shown in \hyperref[语义恢复机制消融对比]{Table \ref*{语义恢复机制消融对比}}, removing the Semantic Recovery (SR) module results in a decrease in precision from 97\% to 89.4\%, lower than Pluto, with an increase of 34 false positives. This significant change fully demonstrates the important role of semantic recovery in enhancing detection precision. Without semantic recovery, the vulnerability detection process lacks support for a deep understanding of contract behavior, leading to an inability to accurately distinguish between normal behavior and potential vulnerable behavior, thus increasing the risk of misreporting normal code as vulnerabilities.

\begin{table}[pos=h!]
\centering
\caption{Impact of semantic recovery on dataset $D_m$.}
\label{语义恢复机制消融对比}
\begin{tabular}{l|ccc|ccc}
\hline
Approach& \multicolumn{3}{c|}{CrossInspector w/o SR} & \multicolumn{3}{c}{CrossInspector} \\ \hline
 & TP &FP & prec. & TP & FP & prec. \\ \hline
Value & 388 & 46 & 89.40\% & 388 &12 & 97\% \\ \hline
\end{tabular}
\end{table}

\begin{figure}[h!]
  \centering
  \begin{lstlisting}
contract FreezableToken is StandardToken {...
    mapping(address => uint256) balances;
    mapping(address => uint) internal freezingBalance;
    function balanceOf(address _owner) public view returns (uint256 balance) {
        return balances[_owner] + freezingBalance[_owner];}
...}
\end{lstlisting}
  \caption{A false positive example of integer overflow.}
    \label{整数溢出假阳性示例}
\end{figure}

\hyperref[整数溢出假阳性示例]{Fig. \ref*{整数溢出假阳性示例}} illustrates an example of an integer overflow false positive. In this specific example, \textit{balance} is used to represent the token balance of an address, where the addition of \textit{balances} and \textit{freezingBalance} is used to calculate the total balance of an address. uint256 is a commonly used unsigned integer type with a range from 0 to $2^{256} - 1$. Since the total token amount in practical applications is typically far below the upper limit of uint256, the addition of these two values does not actually cause overflow. Through the semantic recovery mechanism, the addition operation of \textit{balances} and \textit{freezingBalance} is accurately recovered as an addition operation of two \textit{Balance mapping}, allowing CrossInspector to understand that this operation is logically safe and thus avoiding incorrectly marking it as an overflow. In contrast, other tools (such as Mythril) incorrectly report this situation as an integer overflow.

\subsubsection{Dependency extraction and analysis}


Another advantage of CrossInspector is its significant enhancement of cross-contract analysis capability through the construction of the SDG. The effectiveness of SDG is primarily reflected in its recall, i.e., the ability to reduce false negatives. In the recall evaluation (\hyperref[召回率表]{Table \ref*{召回率表}}), CrossInspector achieves a high recall of 96.75\%, while Pluto, another tool capable of cross-contract analysis, achieves a recall of 82.6\%. Other tools not specifically designed for cross-contract vulnerabilities achieve a maximum recall of only 60.34\%.

\begin{figure}[h!]
  \centering
  \begin{lstlisting}
contract Test1 {
    uint public seed = 10000;
    function getSeed() public returns (uint) {
        return seed;}
    function setSeed(uint x) public {
        seed = x;}
}
contract Test2 {...
    address[] members;
    uint winner = 0xffffffff;
    Test1 t1;
    function addMember() public {
        members.push(msg.sender);}
    function bet() public {
        while (winner == 0xffffffff) {
            uint index = block.timestamp % (t1.getSeed());
            if (index <= members.length) continue;
            winner = index;}
        transferToWinner(1);}
    function transferToWinner(uint amount) private {
        require(winner != 0xffffffff);
        address(members[winner]).transfer(amount);}
...}
\end{lstlisting}
  \caption{A false negative example of DoS attack.}
    \label{DoS假阴性示例}
\end{figure}

Through manual inspection of all false negative results reported by the other four tools, we found that most false negatives could be avoided by conducting more fine-grained cross-contract analysis. \hyperref[DoS假阴性示例]{Fig. \ref*{DoS假阴性示例}} shows an example of a cross-contract DoS vulnerability involving the functions \textit{Test2.bet} and \textit{Test1.getSeed}. In function \textit{Test2.bet}, if \textit{index} remains greater than \textit{members.length}, it will lead to a DoS attack. This vulnerability was missed by Oyente, Mythril, and Slither because these tools did not perform fine-grained inter-procedural analysis. In contrast, CrossInspector effectively identified this vulnerability based on the constructed SDG.

\begin{figure}[h!]
    \centering
    \includegraphics[width=.44\textwidth]{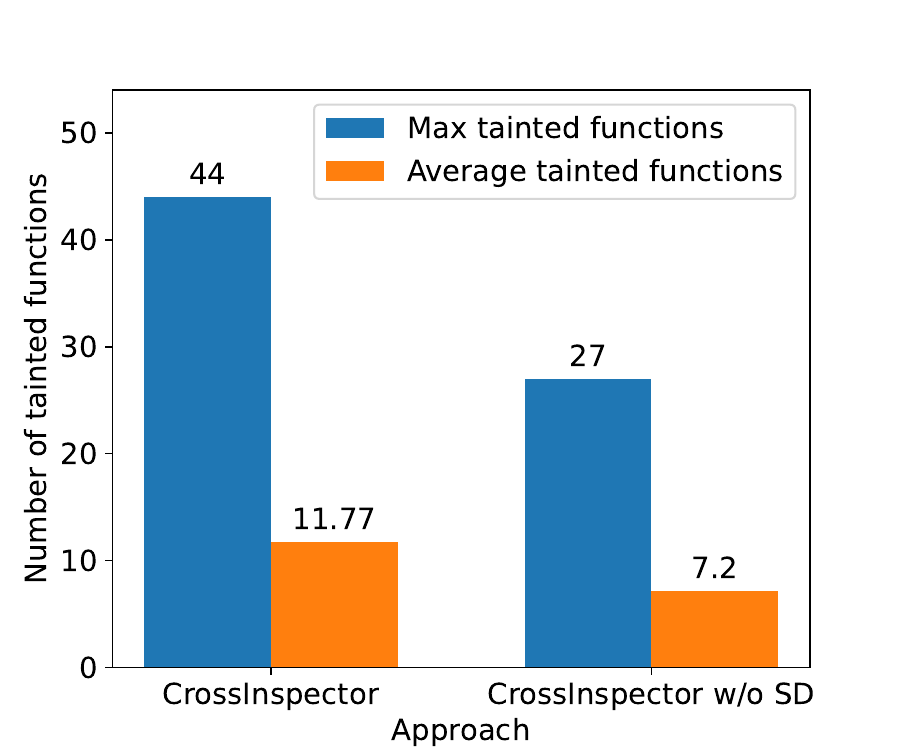}
    \caption{Impact of state dependency edges on dataset $D_m$.}
    \label{RWSD边对受污染函数个数的影响}
\end{figure}

CrossInspector achieves a higher recall compared to Pluto, another tool for cross-contract analysis, primarily due to its more comprehensive analysis of dependencies in smart contracts. CrossInspector not only covers control flow and data flow analysis but also extends to state read-write dependencies and state revert dependencies. As shown in \hyperref[RWSD边对受污染函数个数的影响]{Fig. \ref*{RWSD边对受污染函数个数的影响}}, experimental results demonstrate that without state dependency (SD) edges, the maximum number of tainted functions in a smart contract taint analysis is 27, with an average of 7.20 tainted functions. In contrast, when SD edges are added, the number of tainted functions significantly increases, reaching a maximum of 44 and averaging 11.77 tainted functions. This comparison highlights the role of SD edges in expanding the coverage of taint analysis.

\subsubsection{Parallel memoization path search}

In this section, we investigate the impact of the parallel memoization search mechanism on the efficiency of vulnerability detection. We compare the time overhead of vulnerability detection between using a serial path search strategy and a parallel memoization search strategy.
As shown in \hyperref[并行搜索表]{Table \ref*{并行搜索表}}, when employing the serial path search strategy, the average detection time per smart contract is 16.34 seconds. Upon transitioning to parallel memoization search, this average detection time is reduced to 10.68 seconds, achieving approximately a 1.68 times speed improvement.

\begin{table}[pos=h!]
\centering
\footnotesize
\setlength{\tabcolsep}{3pt} 
\caption{Impact of parallel memoization path search on dataset $D_m$.}
\label{并行搜索表}
\begin{tabular}{lccc}
\hline
     Approach& Serial path search& Parallel path search   & Speedup  \\ \hline
Avg.time(s)    &  16.34 & 10.68      & 1.68x       \\ \hline
\end{tabular}
\end{table}

\subsubsection{Parallel taint analysis}

In this section, we discuss the impact of parallel taint analysis on the efficiency of vulnerability detection. We compare the time overhead of vulnerability detection using serial taint analysis and parallel taint analysis. As shown in \hyperref[并行污点分析机制消融对比]{Table \ref*{并行污点分析机制消融对比}}, experimental results indicate that transitioning taint analysis from traditional serial processing mode to parallel processing mode reduces the average detection time per smart contract from 16.34 seconds to 13.63 seconds, achieving approximately 1.21 times efficiency improvement.

\begin{table}[pos=h!]
\centering
\footnotesize
\setlength{\tabcolsep}{3pt} 
\caption{Impact of parallel taint analysis on dataset $D_m$.}
\label{并行污点分析机制消融对比}
\begin{tabular}{lccc}
\hline
     Approach& Serial taint analysis& Parallel taint analysis   & Speedup  \\ \hline
Avg.time(s)    &  16.34 & 13.63      & 1.21x       \\ \hline
\end{tabular}
\end{table}

However, this improvement is not as significant as the efficiency improvement brought by the previous parallel memoization path search mechanism. This phenomenon mainly arises because the path search phase has already fully explored potential execution paths in the state dependency graph, preprocessing a significant amount of path information for the taint analysis phase. These two phases are actually complementary. When the vulnerability indicator is located near the function entry, parallel taint analysis brings about a greater time improvement; conversely, when the vulnerability indicator is located near the function exit, the time improvement brought by parallel taint analysis is relatively small.

After combining parallel memoization search and parallel taint analysis, the average detection time per smart contract is significantly reduced to 7.83 seconds (\hyperref[时间开销表]{Table \ref*{时间开销表}}). Compared to the non-parallel approach with a detection time of 16.34 seconds, this achieves over 2 times efficiency improvement. Furthermore, considering the time required for the decompilation step itself exceeds 5 seconds, if this time-consuming part is deducted from the total detection time, the detection method based on the two parallel optimization mechanisms actually achieves over 4 times efficiency improvement.

\subsection{Cross-contract vulnerabilities in real world}

To answer RQ4, we conducted detection on dataset $D_r$. Analyzing the detection results of 300 randomly selected smart contracts, we identified 11 cross-contract vulnerabilities missed by the other four tools. We are attempting to reach out to contract developers to ensure these issues are safely addressed.



Case Study. at \textit{0xdd17afae8a3dd1936d1113998900447-ab9aa9bc0}. The code at this address contains cross-contract timestamp manipulation and reentrancy vulnerabilities. Spec-ifically, the function \textit{COIN\_BOX.Collect} relies on the time \textit{now} for subsequent withdrawal logic, which also affects the function \textit{Log.AddMessage} in another contract. This timestamp manipulation vulnerability was overlooked by the other four tools. Additionally, since function \textit{COIN\_BOX.Collect} modifies the balance after the transfer, it may allow attackers to achieve reentrancy attacks. CrossInspector is able to effectively identify both of these vulnerabilities and report the two involved functions as vulnerability paths. This demonstrates the effectiveness of CrossInspector in detecting real-world smart contract vulnerabilities.

\subsection{Discussion and limitation}

CrossInspector has several advantages in cross-contract vulnerability detection: (1).Higher recall and precision are achieved by CrossInspector through effective recovery of contract semantic information and the construction of state dependency graph. (2).CrossInspector improves processor resources utilization and avoids redundant traversal of vulnerability paths, thereby improving detection speed while analyzing more functions and state variables affected by vulnerabilities.

The limitations of CrossInspector mainly manifest in three aspects. Firstly, currently CrossInspector only supports detection of four major types of smart contract vulnerabilities, but it possesses scalability to extend to other types of vulnerability detection as per requirement. Secondly, the decompiler currently employed may lead to inference errors or excessive time consumption in certain cases. Future work could integrate more advanced decompilers to enhance effectiveness and efficiency. Lastly, the semantic recovery model could consider gathering more corpus for training to understand a broader range of real-world contract semantics.

Below, we discuss the soundness and completeness of the mechanisms involved in CrossInspector. The recovery of smart contract semantics is limited by the diversity of the training corpus, as NMT model cannot fully grasp the contextual semantics of smart contracts in the real world, thus introducing unsoundness. Furthermore, both NMT model training and parameter tuning introduce imprecision into NMT predictions, leading to incompleteness. Dependency extraction is limited by the constraints of program logic recovery (i.e., provided by decompiler), resulting in imprecise and incomplete information. The construction of the SDG and vulnerability detection are both complete and sound, without introducing false information or omitting valid information.

\section{Related work}\label{s6}
\setlength{\parindent}{0pt}\textbf{Smart contract vulerability detection.} Smart contract vulnerability detection methods can be generally divided into several categories: static analysis, fuzzing, symbolic execution, and formal verification. Symbolic execution-based tools include Oyente\cite{oyente}, Mythril\cite{mythril}, Maian\cite{maian}, Manticor-e\cite{manticore}, TeEther\cite{teether}, DefectChecker\cite{defectchecker}, Park\cite{park}, and Vand-al\cite{vandal}, etc. Static analysis-based tools comprise SmartChec-k\cite{smartcheck}, SASC\cite{sasc}, and Slither\cite{slither}, SmartAxe\cite{smartaxe}, etc. Fuzzing-based tools encompass ContractFuzzer\cite{contractfuzzer}, Harvey\cite{harvey}, and sFuzz\cite{sfuzz}, etc. Meanwhile, ZEUS\cite{zeus}, VerX\cite{verx}, and Securify\cite{securify}, etc., are grounded in formal verification methods. According to our knowledge, Slither and Mythril are currently the most advanced and popular tools for smart contract vulnerability detection, both of which maintain regular updates on their GitHub repositories. Oyente is the pioneer in the field of smart contract vulnerability detection and continues to be used in many academic evaluations\cite{maian,defectchecker,contractfuzzer,sfuzz,zeus,verx,icse20,reguard,ren2021empirical,so2020verismart,torres2021confuzzius,securify,wang2020contractward,zhang2019soliditycheck,sasc}. In contrast to these tools, Pluto\cite{pluto}, Clairvoyance\cite{clairvoyance}, xFuzz\cite{xfuzz}, and SmartDagger\cite{smartdagger} are specifically designed for cross-contract vulnerabilities. Due to Clairvoyance and xFuzz only supporting source code and Clairvoyance being limited to detecting reentrancy vulnerabilities, they are not versatile enough. Smartdagger is not open-source. Therefore, we chose Pluto for experimental comparison.
\section{Conclusion}\label{s7}
In this paper, we propose CrossInspector, a static analysis-based framework for cross-contract vulnerability detection. We consider a more comprehensive set of smart contract semantics and design a Transformer-based model to effectively recover semantic information from smart contracts to reduce false positives. By extracting control flow and data flow information, as well as state read-write dependencies and state revert dependencies, we construct a state dependency graph to reduce false negatives. We accelerate vulnerability path discovery through function call graph pruning and parallel memoization search techniques, as well as accelerate taint propagation through parallel taint analysis techniques.
The experimental results indicate that CrossInspector surpasses state-of-the-art tools in effectiveness and achieves efficiency on par with the fastest tool that utilizes bytecode for detection.


\bibliographystyle{elsarticle-num}

\bibliography{cas-refs}


\end{document}